# Inertial fusion without compression does not work either with or without nanoplasmonics

*I. B. Földes[1] and G. I. Pokol[2]*

[1]Wigner Research Centre for Physics, Budapest, Hungary

[2]Institute of Nuclear Techniques, Budapest University of Technology and Economics, Budapest, Hungary

**Abstract**

A recently published scheme for inertial fusion based on instantaneous heating of an uncompressed fuel is criticized. It is shown that efficient fusion and "time-like" fusion burn propagation cannot be realized due to the low nuclear reaction cross-sections. The suggested use of nanospheres inside the volume of the target to support the fast heating of the fuel is also questioned.

Since the demonstration of the first laser 60 years ago by Maiman (1960), one of the main motivation of developing more and more energetic lasers is the possible application for pro- ducing nuclear fusion energy by inertial confinement. Inertial confinement fusion was the main driving force of building more and more energetic lasers in the next half century. However, in the last decade, these initiations suffer a serious setback because the largest laser facility, the National Ignition Facility (NIF), could not produce the promised positive energy balance (Brumfiel, 2012). Since then, a slower development started. Hurricane *et al.* (2014) reached more fusion energy than the kinetic energy of the fuel, demonstrating the effect of α-particle heating. Then, further improvements with low-density gas-filled hohlraums resulted in 56 kJ fusion energy. This is at the limit of "burning plasma", a condition where the α-particle deposition is the dominant source of plasma heating as discussed by Hurricane *et al.* (2019). The efficiency of energy coupling into the fuel from the hohlraum is low; therefore in the case of indirectly driven inertial fusion, laser energy as high as 3.5– 4 MJ will be needed to obtain a gain higher than 1 (Kline *et al.*, 2019). Therefore, a number of efforts aim to find alternative ways from which the direct drive scheme seems to have the most perspectives with either a fast ignition or shock ignition scheme. Particular interest is directed toward nonthermal processes in order to reduce the requirements by the $\rho R$ constraints of the Lawson criterion (Lawson, 1957; Kidder, 1976).

In such a situation, all new ideas are welcome, which carry the promise of providing energy gain and still respect the laws of physics. However, recently some new ideas appeared which either do not provide fusion energy gain and/or are in contradiction with the physical possibilities. We feel obliged to show the shortcomings of such ideas in order to mitigate the con- sequences of false hopes.

The ignition scheme formulated by Csernai and Strottman (2015) is claimed by the authors to hold the potential of a breakthrough. Using relativistic hydrodynamics, they calculated the propagation of a detonation wave in an uncompressed sphere for which they obtain the possibility of "time-like propagation", which means a detonation wave propagating with the speed of light. Then, with a logical jump, they claimed that this was a possible scheme for inertial fusion avoiding the Rayleigh–Taylor instability, which had prevented NIF to achieve high gain from compressed fusion pellets. Normally, detonation waves are sorts of shock waves which are fed in the shock front by the chemical reaction (Landau and Lifshitz, 1986). The chemical reaction can, in principle, be substituted by a nuclear reaction. The problem of the paper is that it does not consider a concrete nuclear reaction which must have a cross- section high enough to feed the propagation with nearly the speed of light. Clearly, the greatest problem with this concept is the nuclear fusion cross-sections (e.g., DT) being too low; in fact, several orders of magnitude lower than the cross-sections of the Coulomb collisions. The paper of Csernai and Strottman (2015), however, does not consider this issue at all. Due to the low cross-section of the DT nuclear reaction, the fusion reactions take a long time when the target is not compressed. The target will expand even if it is heated instantaneously; thus, the density decreases further, reducing the efficiency due to the decreasing $\rho R$, which indicates extremely low burn fraction.

The recent paper of Csernai *et al*. (2018) commented here is even more problematic in this respect. The authors carried out calculations for an uncompressed ball heated by external radi- ation. The heating is described by simple geometric consideration with a spherical target. Although the geometric equations may be correct, the conclusions, especially those for inertial fusion, are unfortunately incorrect.

Herewith the basic errors of the paper are summarized:

1. The underlying assumption of the calculations is that the sphere will not be compressed and it will not expand, which is wrong. Even if one heats up the target practically instantaneously, it will immediately start to expand. The cross-section of the DT reactions is low; therefore, relatively long time is needed for fusion reactions; consequently, the efficiency – as determined by the initial density and the time of expansion – is low. This is a well-known phenomenon resulting in an extremely low burn fraction of the fusion fuel, as discussed in the first pages of the book of Lindl (1998), which can also be understood from the book of Atzeni and Meyer-ter-Vehn (2004). According to these textbooks, the fusion efficiency (determined by the fusion cross-section and the thermal expansion) is given by a formula: $w = \rho R/(\rho R + 6 \text{ g/cm}^2)$. In case of a target of initially 1.062 g/cm$^3$ density with a radius of 640 μm – as proposed in the paper by Csernai *et al*. (2018) – $\rho R \approx 0.07$ g/cm$^2$ which results in a burn fraction as low as ≈1%.
2. They aim to heat up the target of 640 μm radius and of 1.062 g/cm$^3$ density to fusion temperature with a 2 MJ laser. In order to heat up a DT plasma to more than 20 keV fusion temperature (as claimed in the paper), $2.3 \times 10^9$ J/g energy is needed (assuming equal electron and ion temperature) which corresponds to somewhat higher, 2.6 MJ laser energy. Unfortunately, for this scheme, α-particle heating cannot be taken into account (Lindl, 1998). Only an approximately 100% efficiency results in a high gain as 100 times, but – as described in the first point – at low density hardly any gain is to be expected. No evidence is shown for the suggested 70–90 keV temperature to be reached in the detonation wave. This would need an avalanche-type process, but without α-particle heating a pumping energy of four times more laser energy would be needed for reaching this optimal temperature, which means more than 10 MJ required laser energy.
3. Some numerical values are confused in the paper, even the intensity considerations are totally erroneous, especially in p. 3 of the paper [see the paragraph: "The average intensity of thermal radiation reaching the surface of the pellet amounts to $Q$ per unit surface (μm$^2$) and unit time (μm/c)…"]. The scheme there requires a laser of 2 MJ energy (as NIF) with a pulse duration of 10 ps. This indicates a thousand times higher power than in NIF, for which 30 time larger optics would be needed. The derived intensity is wrong: In case of the above-mentioned 640 μm target radius (and not 0.64 μm which is probably a typo), the intensity will be not $3.87 \times 10^{20}$ but $3.87 \times 10^{18}$ W/cm$^2$, which is only weekly relativistic, for which neither a relativistic shock wave nor any hole-boring effect is expected.
4. The absorption coefficient is mistreated: It seems that the authors use it as a free parameter, not the temperature- and density-dependent opacity which is a material property. In reality, the opacity of the cold matter is fundamentally different from that of the heated matter. Thus, the visible radiation will not penetrate into the depth of the target, but it will be absorbed within the underdense plasma and the skin layer of the generated plasma. Csernai *et al*. (2018) assume a constant opacity during the whole process, and they do not take into account the variation of opacity during the heating process and hot plasma formation.
5. Another serious problem is that the authors sometimes (especially when discussing variable absorptivity) assume the opacity for X-ray radiation of 1–10 keV failing, however, to mention the way to generate these high-intensity X rays of 10 ps duration. This is actually mentioned in the paper in Chapter 2: "If we want to absorb the whole energy of the incoming laser light on ∼1.3 mm length, we need an absorptivity of $\alpha K \approx 8$ cm$^{-1}$. This is about the absorptivity of DT fuel for soft X-ray radiation of 1 nm wavelength. Longer wavelength radiation would have a larger absorptivity, and would be absorbed in the outside layers of the pellet". Then, the calculations are carried out with this, constant absorptivity, that is, using X-ray absorptivity instead that of the laser light.
6. Csernai *et al.* (2018) mention the possible advantages of the variable absorptivity, which is discussed in Chapter 4. Chapter 5 is the one dealing with the absorption increase using nanospheres, in which several papers are cited about the nanospheres increasing the absorptivity for visible radiation in cold matter. The variable absorption is to be obtained by the variable density of nanospheres, and again visible radiation is mentioned to be used. One can read that "We can experiment with variable absorptivity, which is the normal high temperature, high frequency absorptivity of the DT fuel, $\alpha_{k0} \approx 1$ cm$^{-1}$ at the outer edge of the pellet (i.e., at R = 640 μm) while in the center, it is $\alpha_{ns} = 20–30$ cm$^{-1}$ …" Here again, the whole discussion relies on the assumption that the absorptivity is constant in time. However, as soon as the outer part of the target is heated up and ionized, the opacity changes and the free electrons determine the opacity for visible radiation and the visible radiation cannot penetrate into the target any more. For the absorption of visible radiation, it does not matter whether there are nanospheres inside the volume or not, as the interaction with the visible radiation and its absorption will occur only near the surface of the target, in the underdense plasma. Therefore, the nanostructures inside the target are of no use at all.

Summarizing our comments, it can be concluded that the paper of Csernai *et al*. (2018) starts from unrealistic premises by neglecting the cross-sections of DT reactions, and it is continued by neglecting basic plasma physics, especially by the suggested application of nanospheres in a volume where the laser radiation cannot even penetrate. Thus, the authors of said paper misjudged the possibility of the application of the uncompressed inertial fusion scheme.